\documentstyle[tighten,preprint,eqsecnum,aps,prd]{revtex}
\input{psfig}
\def\no{\nonumber}
\def\a{\alpha}
\def\b{\beta}
\def\e{\epsilon}
\def\g{\gamma}
\def\p{\phi}
\def\t{\theta}

\def\be{\begin{equation}}
\def\ee{\end{equation}}
\def\bi{\begin{itemize}}
\def\ei{\end{itemize}}
\def\cross{\times}
\def\d{\delta}

\begin{document}
\draft

\preprint{\vbox{\baselineskip=12pt
\rightline{IUCAA-25/99}
\rightline{}
\rightline{gr-qc/9906064}
}}
\title{Detection of gravitational waves using a network of detectors
\footnote{Based on talk given at Workshop on Cosmology: Observations confront 
theories, IIT-Kharagpur, India, January 1999. To appear in the workshop 
proceedings being published by Pramana Journal of Physics (The Indian Academy 
of Sciences).}}
\author{Sukanta Bose\footnote{Electronic address:
{\em sbose@iucaa.ernet.in}}, Sanjeev V. Dhurandhar\footnote{Electronic
address: {\em sanjeev@aei-potsdam.mpg.de}}, and Archana Pai\footnote{Electronic
address: {\em apai@iucaa.ernet.in}}}
\address{Inter-University Centre for Astronomy and Astrophysics, Post Bag 4,
Ganeshkhind,\\ Pune 411007, India}

\maketitle

\begin{abstract}
We formulate the data analysis problem for the detection of the Newtonian 
coalescing-binary signal by a network of laser interferometric gravitational 
wave detectors that have arbitrary orientations, but are located at the same 
site. We use the maximum likelihood method for optimizing the detection 
problem. We show that for networks comprising of up to three detectors, 
the optimal statistic is essentially the magnitude of the network correlation 
vector constructed from the matched network-filter. Alternatively, it 
is simply a linear combination of the signal-to-noise ratios 
of the individual detectors. This statistic, therefore, can be interpreted as 
the signal-to-noise ratio of the network. The overall sensitivity of the 
network is shown to increase roughly as the 
square-root of the number of detectors in the network. We further show that 
these results continue to hold even for the restricted post-Newtonian filters. 
Finally, our formalism is general enough to be extended, in a straightforward 
way, to address the problem of detection of such waves from other sources by
some other types of detectors, e.g., bars or spheres, or even by networks of 
spatially well-separated detectors.
\end{abstract}


\section{Introduction}
\label{sec:intro}

The existence of gravitational waves (GW), which is predicted in the theory of 
general relativity, has long been verified `indirectly' through the 
observations of Hulse and Taylor \cite{HT1}. The inspiral of the members of the
binary pulsar system named after them has been successfully accounted for in 
terms of the back reaction due to the radiated gravitational waves 
\cite{HT1,HT2}. However, detection of such waves with man-made `antennas' has 
not been confirmed so far. Nevertheless, this problem has received a lot of 
attention this decade, especially, due to arrival of laser-interferometric 
detectors, which are touted to have the sensitivity required for detecting such
waves.

In the past, a sizable amount of research has been done on the problem of 
detecting gravitational waves using a single bar or interferometric detector. 
However, very little work has been devoted on developing techniques to 
optimally analyze the data from a network of such detectors to seek the 
presence of coalescing binary signal. As has been argued
in the past (see, eg., Ref \cite{BFS}), for a given false-alarm probability, 
the threshold for detection is lowered as the number of detectors is increased.
This increases the probability of detection by a network rather than a single 
detector, provided the observer accepts only coincidences. 
One of the early papers which came close to discussing the problem of detection
of these waves using a network was that of Finn and Chernoff \cite{FC}. This 
paper observed that since the orientations of the two LIGO detectors were very 
similar, their joint sensitivity was larger than any one of them.  
Another work which dealt with the issue of detection using a network was 
that of Bhawal and Dhurandhar \cite{BD}. The main aim of this paper was to 
find the optimal recycling mode of operation of the planned laser 
interferometric detectors for which a meaningful coincidence detection of 
broadband signals could be performed. However, none of these earlier papers
addressed the issue of how a network of detectors with arbitrary 
orientations can be optimally used as a ``single" detector of sensitivity 
higher than that of any of its subsets of individual detectors. One 
of our main aims is to show precisely how this can be achieved. In the process,
we will arrive at a network statistic based on the individual detector outputs
that can be used to ascertain the presence of a signal in them with a given 
level of confidence.

We note that the use of a network has nevertheless received considerable 
attention in the context of the parameter estimation problem. Some of the 
notable works that address this issue are Refs. \cite{DT1,DT2,GT,JK1,JK2}.
The prime motivation in the use of networks in this regard is that the 
larger the number of detectors, smaller the errors in estimated values of 
the binary parameters. However, the starting point in these approaches is 
the assumption that the problem  of detection has already been addressed and 
the detector specific chirp filters that result in ``super-threshold" 
cross-correlations with the individual detector outputs, have been picked.

Here, we formulate the data analysis problem in the case of the 
coalescing binary signal for a network of, say, $N$ number of laser 
interferometric gravitational wave detectors that have arbitrary orientations,
but are located at the same site. The noise in each detector is assumed to be 
additive and Gaussian. Also, the noises in different detectors are taken to be 
independent of one another. We use the maximum likelihood method for optimizing
the detection problem. 

The paper is organized as follows. In Sec. \ref{sec:prelims}, we define the 
various coordinate frames, such as the detector frame and the wave frame, that
we use in our calculations. We describe some known representations of the 
Newtonian signal corresponding to gravitational waves from a coalescing binary.
In Sec. \ref{sec:newsig}, we present a new representation for this signal in 
terms of the complex expansion coefficients of the wave and the detector tensor
in a basis of STF tensors of rank $2$. Section \ref{sec:detect} shows how the 
detection problem can be optimally addressed using the maximum likelihood 
method. In Sec. \ref{sec:sens}, we present the analysis for the improvement of 
the sensitivity of a network as a function of $N$. Finally, in Sec. 
\ref{sec:concl}, we discuss how our results continue to hold for the 
restricted post-Newtonian waveforms. We also mention how our formalism can 
be extended to address the detection problem for a network of spatially well
separated detectors.

We use the following convention for symbols in this paper. Variables 
characterizing the network are displayed in the Sans Serif font. A parenthetic 
index in the superscript or subscript of a variable identifies a particular
detector.  Network- or individual detector-based variables that are 
complex are denoted by uppercase letters, whereas the lower case
letters are reserved for real variables. Note that quantities such as the 
gravitational constant, $G$, though written in upper case, are not complex 
since they do not represent any inherent characteristic of the network or an
individual detector. Also, we define the complex inner product as
$\langle A \Delta_{(I)} ,\> B {\sf f} \rangle = A^* B\langle \Delta_{(I)},\> 
{\sf f} \rangle$. By our convention, all the quantities featuring in the above 
expression, except ${\sf f}$, are complex. Moreover, $\Delta_{(I)}$ denotes a
variable that characterizes the $I$-th detector in the network, where $I$ is a
natural number. Also, ${\sf f}$ denotes a network-based variable.

\section{Preliminaries}
\label{sec:prelims}

We describe the various coordinate frames in terms of which we will analyze 
the different polarizations of an incoming wave. Let $(X,Y,Z)$ be the
orthogonal Cartesian coordinates connected with a weak plane gravitational
wave traveling along the positive $Z$-direction; $X$ and $Y$ 
denote the axes of the polarization ellipse of the wave. Let $(x,y,z)$ 
form a right-handed coordinate system that describes a fiducial detector 
(henceforth referred to as the ``fide'' or the ``network frame''). Let 
us define the Euler angles $\theta$ and $\phi$ to give the incoming direction 
of the wave, and $\psi$ to denote the angle between one semi-axis of the 
ellipse of polarization and the node direction. 
The orthogonal matrix transformation from the wave frame to the fide is thus
defined by the Euler angles $\{\phi,\t,\psi\}$. The 
orthogonal matrix transformation from the fide to the frame of the $I$-th
detector is defined by the Euler angles $\{\a_{(I)},\b_{(I)},\g_{(I)}\}$.


A gravitational wave is represented by metric tensor fluctuation, 
$h_{ij}$, about the vacuum. In the transverse trace-free gauge, its 
non-vanishing components in the wave-frame are $h_{xx} = - h_{yy}{\equiv} h_+ 
$, $h_{xy} = h_{yx} {\equiv} h_{\cross}$). Here, $h_+$ and $h_\cross $ are the 
two polarizations of the waveform. In the Newtonian approximation, they are:
\begin{mathletters}%
\label{h}
\begin{eqnarray} 
{h_+}(t)&=&{2{\cal N} a^{-1/4}(t)\over r} {1+\cos^2 \e \over 2} \cos[\chi(t) + 
\d]\ \ ,\label{hplus} \\
{h_\times}(t)&=&{2{\cal N} a^{-1/4}(t)\over r} \cos\e \sin [\chi(t) + \d]
 \,. \label{hcross}
\end{eqnarray}
\end{mathletters}%
Above, ${\cal N}\equiv \left[ { 2G^{5/3} {\cal M}^{5/3} ({\pi}f_a)^{2/3}/ c^4} 
\right]$,
$r$ is the luminosity distance from the earth to the binary, ${\cal M}$ is
the ``chirp'' mass defined by ${\cal M} = (1+z) \mu^{3/5} m^{2/5}$, where
$m = m_1 + m_2$ is the total mass of the binary, $\mu$ is the reduced mass,
$z$ is the cosmological redshift of the binary, and $c$ is the speed of light 
in vacuum. The angle $\e$ is the angle of inclination of the binary, i.e., the 
angle between the line of sight and the vector normal to the orbit of the 
binary, and $\d$ is an initial phase of the orbital motion. The frequency of 
the gravitational wave is twice the orbital frequency and is given by
\be
\label{ft}
f(t;t_a, {\cal M}) = \left( c^3 \over G \right)^{5/8} {1\over \pi}
\left( {5\over 256 \>{\cal M}^{5/3} \left[t_c (t_a, {\cal M}) - t\right]}
\right)^{3/8} \ \ ,
\ee
where $t_a$ is the time of arrival of the signal (such that $f(t_a )
\equiv f_a = 10Hz$) and $t_c$ is the time at which coalescence occurs. 
Inverting the above equation after setting $f(t_a ;t_a , 
{\cal M}) = f_a$, we get the time of coalescence:
\be
\label{tc}
t_c (t_a , {\cal M}) = t_a + 
{5\over 256 \>{\cal M}^{5/3} (\pi f_a)^{8/3}}\left( {c^3 \over G}\right)^{5/3}
\,.
\ee
Finally, 
$\chi(t) \equiv 2\pi \int_{t_a}^t f(t') dt'$ and
$a(t) \equiv 1 - (t-t_a )/ \xi$,
where 
$\xi = 3.00 \left({{\cal M} / M_\odot }\right)^{-5/3} \left({f_a / 100
\>{\rm Hz}} \right)^{-8/3} \> \> {\rm sec}$,
is the chirp parameter. Note that a total of eight independent 
parameters, viz., $\{r, \d, \t, \p, \psi,\e, t_a, \xi\}$ are required to 
specify this signal. The ranges of the four angles are as follows: $\t\in 
(0,\pi)$, $\p\in (0,2\pi)$, $\psi\in (0,2\pi)$, and $\e\in (0,\pi)$.
 
It can be shown that the signal at the fiducial detector is \cite{SD}
\be \label{sigSD}
s (t) = 2\kappa a(t)^{-1/4} \cos\left(\nu_a \xi 
\left( 1-a(t)^{5/8} \right) + \d - \eta \right) \ \ ,
\ee
where 
\begin{mathletters}%
\label{nu-kappa-zeta}
\begin{eqnarray}
\nu_a = 320\pi \left({f_a \over 100\> {\rm Hz} }\right)\> {\rm Hz} 
\>,&\quad& \kappa \equiv {\cal N}\zeta /r \ \ , \label{kappa} \\
\zeta (\e) \equiv  \left[ {( 1+ \cos^2 {\e})^2 \over 4} + 
\cos^2 {\e} \right]^ {1/ 2} \>,&\quad&
\tan\eta = {2{\cos \e}\over  1+{\cos^2\e}} \ \ , \label{eta}
\end{eqnarray}
\end{mathletters}%
with $\eta \in ( -\pi/4,\> \pi/4 )$. The signal at the $I$-th detector 
can be expressed in terms of the quantities defined above as
\be \label{primisig}
s_{(I)} (t) = o_{(I)+} h_{(I)+} (t) + o_{(I)\times} h_{(I)\times} (t) \,.
\ee
where $h_{(I)+}$ and $h_{(I)\times}$ are the two polarizations of the wave 
arriving at the $I$-th detector. Also, $o_{(I)+}$ and $o_{(I)\times}$ are the
beam pattern functions, which depend on $\{\p, \t, \psi\}$ and the orientation
angles $\{\a_{(I)},\b_{(I)},\g_{(I)} \}$. The problem with the above 
representation for the signal is that it mixes up the factors dependent on the 
detector specific Euler angles $\{\a_{(I)},\b_{(I)}, \g_{(I)} \}$ and those 
dependent on the angles $\{\p, \t, \psi\}$. We now give a different 
representation of the signal where this problem does not occur. It 
will prove useful to address the detection problem for a network.

Consider the complex null vector ${\bf M} \equiv 1/\sqrt{2} ({\bf e}_X +i {\bf 
e}_Y )$, where ${\bf e}_X$ and ${\bf e}_Y$ are real unit vectors in the $X$ 
and $Y$ directions, respectively, of the wave-frame. Then the wave tensor 
$w_{ij}$ is defined as
\be \label{wavetensor}
w_{ij} = h_+ {\rm Re} (M_i M_j) + h_\times {\rm Im} (M_i M_j) \ \ ,
\ee
which is a real symmetric trace-free (STF) tensor. The components of ${\bf M}$ 
in the detector axis are \cite{DT1}:
${\bf M} =  1/\sqrt{2} (\cos\p -i\cos\t\sin\p,\>  \sin\p +i\cos\t\cos\p, \>
i\sin\t ) \exp(-i\psi ) $.
A detector can also be represented as an STF tensor. For an interferometer
with arms along the directions ${\bf n}_1$ and ${\bf n}_2$ (both being unit 
vectors), the detector tensor ${\bf d}$ is given by
\be \label{detectensor}
d_{ij} = n_{1i}n_{1j} - n_{2i} n_{2j} \,.
\ee
The response amplitude of the detector or, equivalently, the signal is 
just the scalar product of the wave and detector STF tensors,
\be \label{sigwd}
s = w^{ij} d_{ij}\ \ ,
\ee
where it is implicit that the Einstein summation convention holds over the 
repeated upper and lower indices $i$ and $j$. In the following analysis, we 
will specifically consider a network of interferometric detectors. However, 
the generalization to bar detectors is straightforward.

Since we extensively deal with STF tensors of rank 2, enunciating some 
frequently used properties of such objects is in order. \footnote{For a 
detailed discussion, see Refs. \cite{GMS,KT1}. For a more selective reading 
for immediate use, we refer to Ref. \cite{DT1}.} Since such tensors have five 
independent elements, they can be expanded in terms of (location-independent)
``STF-$l$'' tensors, ${\cal Y}_{lm}^{ij}$, with rank $l=2$. They are related to
the spherical harmonics as follows:
\be \label{stf2}
Y_{2m} (\t,\p) = {\cal Y}_{2m}^{ij} n_i n_j \>, \quad {\rm where} \quad
m=\pm 2, \>\pm 1,\> 0\ \ ,
\ee
where ${\bf n} = (\cos\p\sin\t,\> \sin\p\sin\t,\> \cos\t )$. There are five
independent $3\times3$ complex matrices, ${\cal Y}_{2m}^{ij}$, obeying the 
normalization
\be\label{norm-stf2}
{\cal Y}_{2m}^{ij} {\cal Y}^{2m'*}_{ij} = {15\over 8\pi}\d_m^{m'} \,.
\ee
In this paper, we will also be interested in the behavior of these STF tensors 
under action of an element $g(\a,\b,\g)$ of the rotation group SO(3), where 
$(\a,\b,\g)$ are the Euler angles. Consequently, we mention that under
such an action, the spherical harmonics obey the following transformation law:
\be \label{transY}
Y_{2m'} (\t ',\p ') = \sum_{m = -2}^2 T_{-m'}{}^{-m} (\a,\b,\g) Y_{2m} (\t,\p)
\ \ , \ee
where $T_{-m'}{}^{-m} $ are the Gel'fand functions of rank 2 \cite{GMS,DT1}. 

As shown in Ref. \cite{DT1}, in the detector frame the wave tensor $w_{ij}$ 
can be expanded in terms of the STF-2 tensors as:
\be \label{wstf2}
w^{ij}(t) = {\sqrt{ 2\pi\over 15}} \left[ h_+(t)(T_2{}^n + T_{-2}{}^n)- i
h_\times (t) (T_2{}^n - T_{-2}{}^n)\right] {\cal Y}^{ij}_{2n} \ \ ,
\ee
where the expansion coefficients are combinations of the Gel'fand functions,
which depend on the parameters $(\p,\t,\psi)$. For interferometric detectors
with arms making an angle of $2\Omega$, the only non-vanishing detector-tensor 
components in its own frame are $d_{12} = d_{21} = \sin 2\Omega$. The 
following analysis, where we will deal with the case $\Omega =\pi/4$, can be
easily generalized to other values of $\Omega$.

Consider the $I$-th detector of a network. 
Using the wave- and detector-tensor components in Eq. (\ref{sigwd}), 
it was shown in Ref. \cite{DT1} that the signal takes the form:
\begin{equation}
\label{signalDT}
s_{(I)} (t) =  I^m_{(I)n} 
T_m{}^s (\p, \t,\psi ) {T_{(I)s}}^{n*} (\a_{(I)} , \b_{(I)} , \g_{(I)} ) 
\ \ , 
\ee
Note that $T_{(I)m}{}^n$ is to be distinguished 
from $T_m{}^n$ in that $T_m{}^n = T_m{}^n (\p,\t, \psi)$, whereas $T_{(I)m}{}^n
= T_m{}^n ( \a_{(I)},\b_{(I)}, \g_{(I)} )$. Above,
\begin{mathletters}%
\label{Imn}
\begin{eqnarray} 
I^m_{(I)n} &=& \pm {i\over 2} ( h_{+(I)} -ih_{\cross(I)} ) \equiv \pm J_{(I)}
\qquad {\rm for} \qquad m=2 \>, \quad n=\pm 2 \ \ ,\label{Imn+} \\
I^m_{(I)n} &=& \pm {i\over 2} ( h_{+(I)} +ih_{\cross(I)} ) = \mp J_{(I)}^* 
\qquad {\rm for} \qquad m=-2 \>, \quad n=\pm 2 \,. \label{Imn-}
\end{eqnarray}
\end{mathletters}%
The signal given in (\ref{signalDT}) has the advantage of keeping factors 
dependent on the two sets of Euler angles separate.

\section{A new representation for the signal}
\label{sec:newsig}

To address the detection problem, we will be directly dealing with a 
construct known as the likelihood ratio (LR) (to be defined in Sec. 
\ref{sec:detect}). The LR is a non-linear functional of the signal. To keep 
this functional form simple we develop a new representation for the signal 
based on (\ref{signalDT}).

In terms of $J_{(I)}$, the signal given in
Eq. (\ref{signalDT}) takes the form
\be \label{sigJG}
s_{(I)} = 2{\rm Re} (iJ_{(I)}^* \Gamma_{(I)} ) \ \ ,
\ee
where
${\Gamma}_{(I)} = i \left(T^{2*}_{(I)s} - T^{-2*}_{(I)\>s} \right)T_{-2}{}^s$. 
Next, we define:
\begin{eqnarray} 
h_{c(I)} &=& a_{(I)}^{-1/4} (t){\cos\chi_{(I)}(t)} \equiv g_{(I)} s_{(I)0}
\ \ , \no\\ 
h_{s(I)} &=& a_{(I)}^{-1/4} (t) {\sin\chi_{(I)}(t)}\equiv g_{(I)} 
s_{(I)\pi /2} \,. \label{hcs}
\end{eqnarray} 
where $g_{(I)}$ is the maximum signal-to-noise ratio for the $I$-th detector
obtainable by using an optimal filter:
\be \label{gI}
g_{(I)}^2 \equiv \int df { |\tilde{h}_{c, s} (f) |^2 \over s_h^{(I)} (f)}\ \ ,
\ee
$s_h^{(I)} (f)$ being the noise power-spectral-density (p.s.d.) of the $I$-th 
detector. Above, $s_{(I)0}$ and $s_{(I)\pi /2}$ can be represented as the real 
and imaginary parts of a complex quantity:
\be \label{SI}
S_{(I)} (t) = s_{(I)0} + i s_{(I)\pi /2} \ \ ,
\ee
which has a norm equal to 2.

Let us now express $\Gamma_{(I)} = \gamma_{(I)0} + i\gamma_{(I)\pi/2}$,
where $\gamma_{(I)0}$ and $\gamma_{(I)\pi/2}$ are, respectively, the real and
imaginary parts of $\Gamma_{(I)}$. Define
\be
\label{W}
W_{(I)} \equiv g_{(I)} \left( \gamma_{(I)0} \cos \eta -i\gamma_{(I)\pi/2}
\sin\eta \right) \equiv w_{(I)} e^{-i\omega_{(I)}} \ \ ,
\ee
where the last expression is the polar form of $W_{(I)}$. Note that the only
signal parameters on which $W_{(I)}$ depends are $\{\p,\t,\psi,\eta\}$. Armed 
with these definitions, the signal in Eq. (\ref{sigJG}) can be re-expressed as
\be \label{sigW}
s_{(I)} (t) = 2 \kappa {\rm Re} \left( W_{(I)}^* R_{(I)} 
\right) \ \ ,
\ee
where we have defined
\be \label{RI}
R_{(I)} (t) = r_{(I)0} + i r_{(I)\pi /2} \equiv S_{(I)} e^{i\delta} \ \ ,
\ee
which is just the rotated $S_{(I)}$.
Here $r_{(I)0}$ and $r_{(I)\pi /2}$ are, respectively, the real and imaginary 
parts of $R_{(I)}$. Equation (\ref{sigW}) is a new representation for the 
signal that we will find useful in obtaining the maximum likelihood ratio 
below.

We end this section by arriving at a relation between the complex variable 
$W_{(I)}$ and the detector tensor. First, by using Eqs. (\ref{wstf2}) and
(\ref{RI}) we obtain the inner products between the wave tensor $w^{ij}$ and 
the real and imaginary components of $R_{(I)}$:
\be\label{abij}
2 \kappa g_{(I)} \a^{ij} =\langle w^{ij} (t), r_{(I) 0}\rangle  
\quad {\rm and} \quad
2 \kappa g_{(I)} \b^{ij} =\langle w^{ij}(t), r_{(I)\pi/2}\rangle 
\ee
where we have defined two new STF tensors $\a_{ij}$ and $\b_{ij}$. These are 
real functions of $\{\p,\t ,\psi, \eta\}$. 

By comparing the different 
representations (\ref{sigwd}) and (\ref{sigW}) of the signal we obtain
\be
\label{WIdef}
W_{(I)} =  \Delta^{ij} \left( g_{(I)} d_{(I)ij} \right)\>,\>\>{\rm where}
\>\> \Delta^{ij} \equiv (\a^{ij} + i\b^{ij} ) \ \ ,
\ee
$\Delta^{ij}$ is a complex STF tensor dependent on $\{\p,\t ,\psi, \eta\}$, 
when expressed in the fide frame. Note that 
\be
|W_{(I)}|^2 = |\a^{ij}\left( g_{(I)} d_{(I)ij} \right)|^2 
              +|\b^{ij}\left( g_{(I)} d_{(I)ij} \right)|^2 \,.
\ee
Above, up to an $r$-dependent factor, $|W_{(I)}|^2$ can be interpreted as the 
total power transferred to the $I$-th detector. More appropriately, it is the
gain factor associated with the $I$-th detector. We have resolved it as a sum 
of the fractions of power transferred to the detector by the two polarizations,
respectively. It can be shown that, up to a factor of $g_{(I)}$, $W_{(I)}$ is 
just a direction cosine that is dependent on the set of angles $\{\phi,\t,\psi,
\eta\}$. This is what one would expect from the above interpretation of 
$W_{(I)}$ as a gain factor. 

\section{Addressing the detection problem for a network}
\label{sec:detect}


The signal from a coalescing binary will typically not stand above the 
broadband noise of the interferometric detectors; the concept of an absolutely 
certain detection does not exist in such a case. Only probabilities can be 
assigned to the presence of an expected signal. In the absence of prior 
probabilities, such a situation demands a decision strategy that maximizes the 
detection probability for a given false alarm probability. This is termed as 
the Neyman-Pearson criterion (see, eg., Ref. \cite{Hels}). Such a criterion 
implies that the decision must be based on the value of a statistic called the 
likelihood ratio (LR). It is defined as the ratio of the probability that a 
signal is present in an observation to the probability that it is not.

For a network of detectors we obtain this statistic as follows. We assume that 
the noise at each detector is additive, Gaussian, and both statistically as 
well as algebraically independent of the noise in any other detector in the 
network. Under these conditions, the network LR, denoted by $\lambda$, is 
just a product of the individual detector LR's. Similarly, the logarithmic 
likelihood ratio (LLR), $\ln \lambda$, can  be verified to have the same form 
as for an individual detector, namely,\cite{Hels,JK1}.
\be\label{MLRNW}
\ln\lambda = \langle {\sf s},{\sf x}\rangle_{NW} -
{1 \over 2} \langle {\sf s}, {\sf s}\rangle_{NW} \ \ ,
\ee
where the normalized set of signals are denoted by a single network vector
\be
\label{sNW}
{\sf s} (t) =\left(s_{(1)} (t), s_{(2)}(t),......., s_{(N)} (t) \right) \ \ ,
\ee
$N$ being the number of detectors in the network. The subscript $NW$ denotes 
that the inner product is defined on the {\em network} space. Similarly, the 
individual detector outputs $x_{(I)} (t)$ are combined to form the network 
vector
\be
\label{xNW}
{\sf x} (t) =\left(x_{(1)} (t), x_{(2)}(t),......., x_{(N)} (t) \right) \,.
\ee
Thus, in terms of the individual detector signals, the LLR is
\begin{eqnarray}
\ln\lambda &=& \sum_{I=1}^N \langle s_{(I)}, x_{(I)} \rangle_{(I)} -
{1 \over 2}\sum_{I=1}^N \langle s_{(I)}, s_{(I)} \rangle_{(I)} \no \\
&=& {\sf b} \sum_{I=1}^N \langle z_{(I)}, x_{(I)}\rangle_{(I)} - 
{1\over 2} {\sf b}^2 \ \ , \label{LRb}
\end{eqnarray}
where 
\be \label{bNW}
{\sf b}\equiv 2\kappa \sqrt{\sum_{I=1}^N w_{(I)}^2 } \ \ ,
\ee
is the norm of ${\sf s}$ and $z_{(I)} = s_{(I)}/{\sf b}$. The aim now is to 
maximize the LLR over all eight parameters to obtain the maximum (logarithmic)
likelihood ratio (MLR). It is the MLR that must be compared with a threshold 
value to ascertain the presence or absence of signal in the detector output, 
with a given level of confidence.

We now analytically maximize the above expression with respect to as
many of the eight parameters as possible. 
Note that the luminosity distance 
$r$ appears only through ${\sf b}$ in LLR. Maximizing it with respect to
${\sf b}$ yields 
\be
\ln \lambda |_{\hat{\sf b}} = {1\over 2} \left(\sum_{I=1}^N \langle z_{(I)},
x_{(I)} \rangle_{(I)} \right)^2  
={1\over 2}\left|{\rm Re}\left( e^{-i\d}\sum_{I=1}^N Q_{(I)} C_{(I)}^*
\right)\right|^2
\ \ , \label{LLRhatb}
\ee
where we have defined
\be \label{CI}
Q_{(I)} \equiv {W_{(I)}\over \sqrt{\sum_{I=1}^N w_{(I)}^2}} \quad {\rm and}
\quad C_{(I)}^* \equiv \langle S_{(I)}, x_{(I)}\rangle_{(I)} \,.
\ee
The network vector ${\sf S}$, with the $S_{(I)}$'s as its components, is the
matched network-filter.

Next we maximize the LLR in Eq. (\ref{LLRhatb}) with respect to $\d$ for,
apart from the phase factor, none of the other terms there depend on it. 
This gives 
\be
\label{LLRhatbd}
\ln \lambda |_{\hat{\sf b},\hat{\d}} = {1\over 2}
\left|\sum_{I=1}^N Q_{(I)} C_{(I)}^*\right|^2
\ \ , \ee
which is a function of six parameters, namely, $\{\p,\t,\psi,\eta,t_a,\xi\}$.
Note that when all the detectors are ``closely" located, it is only the
$Q_{(I)}$'s that depend on four angles $\{\p,\t,\psi,\eta\}$; the $C_{(I)}$'s
then depend only on $\{t_a,\xi\}$, with all the times of arrival being equal.
We will refer to this situation as the ``same-site" approximation. When the 
detectors are spatially well separated, the $C_{(I)}$'s will depend on
$\{\p,\t\}$ as well.

To obtain the MLR, we need to maximize over these remaining parameters. At this
stage it is useful to define the surrogate statistic (SS),
$\lambda' \equiv \ln \lambda |_{\hat{\sf b},\hat{\d}}$.
For a network comprising of a total of $N$ detectors located within a fraction 
of a wavelength, the SS is maximum when
${\sf Q} \propto {\sf C}$,
where ${\sf C}$ is the network correlation vector with $C_{(I)}$'s as its 
components.
Therefore, once ${\sf C}$ is known, the maximization procedure determines
${\sf Q}$ through the above condition and the fact that ${\sf Q}$ has a unit
norm. However, the ${\sf Q}$ so determined will, in general, yield an 
overdeterministic set of equations for the four parameters 
$\{\p,\t,\psi,\eta\}$. On the other hand, if this set of equations can be
solved to yield a physically realizable solution for the parameters, then
the maximized LLR will have a simpler form:
\be\label{DLLRhatbd}
\lambda '|_{\hat{\p},\hat{\t},\hat{\psi},\hat{\eta}} 
= {1\over 2}\sum_{I=1}^N c^2_{(I)} \equiv {\Lambda\over 2}
\ \ , \ee
where $c_{(I)}$ is the magnitude of $C_{(I)}$. Above, $\Lambda$ is a 
function of two parameters, namely, $\{t_a,\xi\}$. Although $\Lambda$ is a 
{\em real} quantity, we follow the established convention in literature
to denote the LLR by an uppercase letter! 


It can be shown that the condition ${\sf Q} \propto {\sf C}$ is always realised
for two detectors. Numerical calculations suggest that this result holds for
three detectors as well \cite{BDP}.
However, for networks with a larger number of detectors we numerically find 
that this condition is not always realisable and one is forced to maximize 
$\lambda '|_{\hat{b}\hat{\d}}$, as given in Eq. (\ref{LLRhatbd}), over the 
four angles. Thus, for networks comprised of up to three detectors, the 
application of Eq. (\ref{DLLRhatbd}) appears to be valid. We will limit our
discussion to only such cases below. Hence, only the maximization of $\Lambda$ 
over the two parameters, $\{t_a, \xi\}$, remains to be done. This is performed 
numerically along the lines of Sathyaprakash and Dhurandhar (see Ref. 
\cite{SD}). 

\section{Network sensitivity}
\label{sec:sens}

To infer the presence of a signal from the outputs of the members of a 
network, one compares the value of the statistic $\Lambda$ in Eq. 
(\ref{DLLRhatbd}) with a predetermined detection threshold $\Lambda_0$. As we 
show below, the value of $\Lambda_0$ can be obtained (via the Neyman-Pearson 
decision criterion \cite{Hels}) from the false alarm probability, $Q_0$, 
associated with the event of detection of such a signal. For $\Lambda < 
\Lambda_0$, presence of a signal in the data is ruled out, whereas if $\Lambda 
> \Lambda_0$, then the detection of a signal in the data is announced. 

We now analyze the improvement in the sensitivity of a network over that of
a single detector. Apart from the assumptions about detector noise mentioned in
Sec. \ref{sec:detect}, we further assume that it is stationary, which implies 
that for the $I$-th detector we have
\be
\langle\tilde{n}_{(I)}(f) \tilde{n}^{*}_{(I)}(f) \rangle = s_{h(I)} \delta(f
-f') \ \ ,
\ee
where $s_{h(I)}$ is the noise p.s.d. of that detector and the angular brackets 
imply ensemble average. In general, different detectors may have non-identical 
noise p.s.d. We will assume that the noise in any detector is white, i.e.,  
$s_{h(I)}$ does not vary with frequency. Note that it has zero mean, i.e, 
$\langle\tilde{n}_{(I)}(f)\rangle = 0$, and its standard deviation is 
$s_{h(I)}$. 

Equation (\ref{DLLRhatbd}) shows that $\Lambda$ is a sum of squares of 
independent random variables with Gaussian probability distribution functions
(PDF). Thus, $\Lambda$ itself must have the so-called $\chi^2$ probability 
distribution. Hence, the PDF of $\Lambda$ under the hypothesis that the 
signal is present, $H_1$, is:
\be\label{PDF1Lam}
p_1 (\Lambda) = {1\over 2} \left({\Lambda\over {\sf b}^2}\right)^{(N-1)/2} 
\exp[-(\Lambda+{\sf b}^2)/2]\> I_{N-1} ({\sf b}\sqrt{\Lambda}) \ \ ,
\ee
where $\Lambda > 0$ and $I_{N-1} (x) $ is the modified Bessel function of order
$(N-1)$. Note that ${\sf b}^2$ is proportional to the total energy received 
from the source. When ${\sf b}\sqrt{\Lambda} >> 1$, the above expression 
approximates to
\be \label{PDF1Lamapp}
p_1(\sqrt{\Lambda}) = {1 \over \sqrt{2\pi}}\exp[-(\sqrt{\Lambda}-{\sf b})^2/2] 
\,. \ee 
On the other hand, under the hypothesis that the 
signal is absent, $H_0$, the PDF of $\Lambda$ is
\be \label{PDF0Lam}
p_0 (\Lambda) = {(\Lambda/2)^{N-1} e^{-\Lambda/2} \over 2(N-1)!} \ \ ,
\ee
which one can obtain from Eq. (\ref{PDF1Lam}) by taking the limit ${\sf b}\to 
0$.

\begin{figure}  \label{LthreshA}
\centerline{\psfig{file=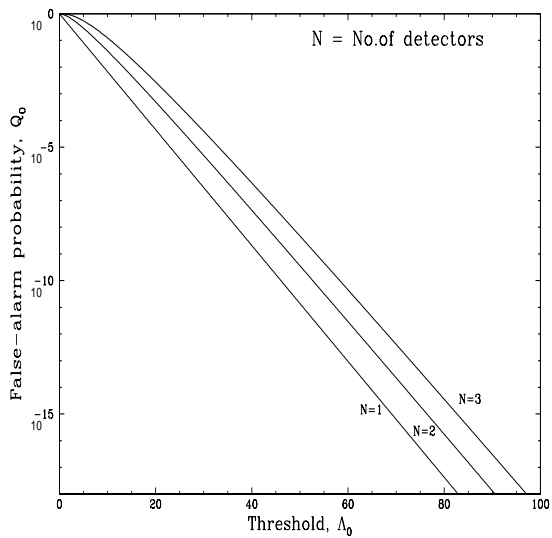,height=3.0in,width=3.0in,angle=270}}
\vskip 0.5in
\caption{The plot of $Q_0$ as a function of the detection threshold $\Lambda_0$
for different number, $N$, of closely located detectors in a network.}
\end{figure}

We are now in a position to calculate the false-alarm probability:
\be\label{Q0La}
Q_0 = \int_{\Lambda_0}^\infty p_0 (\Lambda) d\Lambda =1-I(\Lambda_0 /2\sqrt{N},
 \>N-1) \ \ ,
\ee
where we have made use of the incomplete gamma-function
\be \label{incomgam0}
I(u,\> p) = \int_0^{u\sqrt{p+1}} x^p e^{-x} dx/p! \>\>\,.
\ee
For a given false-alarm probability $Q_0$, Eq. (\ref{Q0La}) allows us to obtain
the detection threshold $\Lambda_0$ of our statistic. Plots of $Q_0$ versus 
$\Lambda_0$ for different values of $N$ are given in Fig. 1. From these plots
it can be inferred that $\Lambda_0$ increases slowly with $N$.

The detection probability $Q_d$ can be obtained by computing the area under the
function $p_1 (\sqrt{\Lambda})$ for $\sqrt{\Lambda} >\sqrt{\Lambda_0}$. 
For ${\sf b}\sqrt{\Lambda} >> 1$, it is 
\be\label{QdLa}
Q_d = \int_{\sqrt{\Lambda_0}}^\infty p_1 (\sqrt{\Lambda}) d({\sqrt{\Lambda}}),
\ee
where $p_1$ is given by Eq. (\ref{PDF1Lamapp}). We now show that for a given 
$Q_d$ and $Q_0$, the distance $r$ up to which a
network can probe increases with $N$. This is tantamount to saying that the 
sensitivity of a network increases as a function of $N$. For simplicity, assume
that the detectors are oriented in such a way that ${\sf b}^2$ is proportional 
to $N$. This is the case when, e.g., the $w_{(I)}$'s are all identical.
Let $Q_d=0.5$, i.e., $\sqrt{\Lambda_0} = {\sf b}$. As $N$ increases, 
$\Lambda_0$ increases for a given $Q_0$ (see Fig. 1.). However, given the fact 
that ${\sf b} \propto \sqrt{N}/r$, we  have $r \propto \sqrt{N}/\Lambda_0$. 

\begin{table} \label{tab:q01b}
\begin{center}  
\caption{The ratio of the sensitivity of a network with $N = 2,3$, 
respectively, relative to a single detector for $Q_d = 0.5$, and corresponding
to different false-alarm probabilities.}
\vspace{0.4in}
\begin{tabular}{|c|c|c|c|c|}  
\multicolumn{4}{c}{\hspace {1.5in} False-alarm probability, $Q_0$} \\ 
\cline{1-5}  
No. of detectors, $N$
&0.33$\times 10^{-10}$
&0.67 $\times 10^{-12}$ &0.17 $\times 10^{-12}$ &0.33 $\times 10^{-14}$ \\
\hline
2 & 1.33 & 1.33 & 1.33 & 1.34 \\ 
\hline
3 & 1.55 & 1.56 & 1.57 & 1.58 \\
\end{tabular} 
\end{center}
\end{table}

\noindent Thus, given a specific binary, the ratio of sensitivity of a
network to that of a single detector behaves as
\be
{r(N)\over r(N=1)} = \sqrt{N}\left(\Lambda_0 (N=1)\over \Lambda_0 (N) 
\right)^{1/2} \ \ ,
\ee
which can be computed by using $\Lambda_0 (N)$ from Fig. 1. These ratios, which
are presented in Table 1. for  $N=2$, $3$, clearly show that the sensitivity of
such a network increases a little slower then $\sqrt{N}$. This implies an 
increase in the survey volume accessible to a network, which, in turn, implies
an increased event rate.

\section{Conclusion}
\label{sec:concl}

The problem of detecting inspiral waveforms from coalescing binaries via 
pattern-matching can be made more accurate by including post-Newtonian 
corrections in the corresponding filters. In this regard, it has been shown 
\cite{CF} that it is both necessary and sufficient to work with the 
{\em restricted} post-Newtonian chirp. The description of the resulting 
waveform involves an extra parameter, apart from the set of eight 
parameters described above. However, as was shown by Sathyaprakash \cite{BS1}, 
for the astrophysically relevant range of parameters, the effective 
dimensionality of the parameter space remains unchanged. Hence, even after
the inclusion of the restricted post-Newtonian corrections in the filters, 
the detection problem can be addressed in the same manner as described in Sec.
\ref{sec:detect}.

When the detectors are spatially well separated, the cross-correlations, 
$C_{(I)}$'s, will be dependent on the times of arrival, $t_{a(I)}$'s, which are
different from one another. Since a specific $t_{a(I)}$ depends on the location
angles $\{\p,\t\}$, so will $C_{(I)}$. Hence the maximization of the SS over 
the four angles $\{\p,\t,\psi,\eta\}$ that was performed for the same-site
approximation above, can no longer be implemented in 
the present case. It can be shown that the SS can be recast in such 
a way that its dependence on the complementary angles $\{\psi, \eta\}$ is 
isolated. This aids in the analytic maximization of the SS over these
two angles. The maximization over the remaining four parameters $\{\p,\t,t_a ,
\xi\}$ can then be effected numerically on a four-dimensional parameter-space
grid. Details of these calculations will be presented elsewhere \cite{BDP}.

\section{Acknowledgments}

One of us (SB) wishes to thank the organizers of the conference for their
kind hospitality. SVD thanks the Department of Physics and Astronomy,  
Cardiff University, U.K., for hospitality. AP thanks the CSIR, India, 
for Senior Research Fellowship.



\end{document}